\documentclass[prb,preprint,showpacs]{revtex4}
\usepackage{epsfig}
\usepackage{color}
\usepackage{ulem}

\begin{document}
\title{Effect of Coulomb correlation on electron transport through
concentric quantum ring-quantum dot structure}

\author{T. Chwiej}
\email{chwiej@novell.ftj.agh.edu.pl}

\author{K. Kutorasi\'nski}

\affiliation{Faculty of Physics and Applied Computer Science,
AGH University of Science and Technology,
al. Mickiewicza 30, 30-059 Krak\'ow, Poland}

\date{\today}

\begin{abstract}

We study transfer of a single-electron through a quantum ring capacitively coupled to the charged
quantum dot placed in its center. For this purpose we solve the time-dependent Schr\"odinger
equation for the pair of particles: the electron traveling through the ring and the other carrier
confined within the quantum dot. The correlation effects due to the interaction between the charge
carriers are described in a numerically exact manner. We find  that the amplitude of Aharonov-Bohm
oscillations of the transfer probability is significantly affected by the presence of the
dot-confined carrier.  In particular the Coulomb correlation leads to inelastic scattering. When the
inelastic scattering is strong the transmission of electron through the ring is not completely
blocked for $(n+1/2)$ magnetic flux quanta.

\end{abstract}

\maketitle

\section{\label{sec:intro} Introduction}

Semiconductor quantum rings allow for observation of the electron self-interference.
When electron traverses the ring threaded by magnetic field it is subject
to a constructive or a destructive interference what manifests itself as conductance oscillations. This effect
as predicted by Aharonov and Bohm\cite{theoryAB_1} was observed in many experiments with quantum
rings\cite{ex_AB_1}. Manipulation of electron wave function phase in both arms of the quantum ring
allows to obtain strong or weak coupling between the ring and the leads by tuning
the magnetic field. Recently, besides the most intensively examined two-terminal open quantum
rings\cite{ex_2_terminal_1,ex_2_terminal_2,ex_2_terminal_3} this kind of coupling was experimentally
tested in three-terminal \cite{3lorentz,ex_3_terminal_2} as well as in four-terminal quantum
rings\cite{ex_4_terminal_1}. The current oscillations are highly sensitive to decoherence resulting from
interaction with environment such as electron-phonon or electron-electron interactions.
	Effect of electrostatic interaction on magnetotransport was observed experimentally for ring
with quantum dot placed in one of its arms\cite{ex_dot_ring_2}, for ring capacitively coupled to the
quantum dot placed beside it\cite{ex_dot_ring_3,ex_dot_ring_1} as well as  for rings working in
Coulomb blockade regime and confining from few\cite{ex_cotunneling} to several hundreds
electrons.\cite{ex_interaction_ring_2,ex_interaction_ring_1} The weak localization theory
predicts  that the phase coherence time against the effects of the  electron-electron interaction approaches infinity for
zero temperture\cite{mohanty}. However, besides the decoherence, which is suppressed in low
temperature, the electrostatic interaction is also responsible for existence of spatial correlations
between charged particles. We may divide correlations induced by electrostatic interaction in a crude way
in two types: (i) the Coulomb correlation which introduces dependence of mutual particle positions
due to repulsive or attractive interaction and (ii) the Pauli correlation which arises directly from
the Pauli exclusion principle.
	An extremely strong effect of Coulomb correlation on magnetotransport in quantum ring was
experimentally observed by M\"uhle et al.\cite{ex_ring_ring_1} Measurements of magnetic field
dependence of conductance for system of two concentric capacitively coupled quantum rings revealed
two-period oscillations which authors ascribed to existence of the AB effects in the inner and in
the external ring.
This experiment explicitly proves that Coulomb correlation may greatly affect electron transport in
quantum ring even in low temperatures when the decoherence due to electron-electron scattering
vanishes.

In this paper we study the single-electron transport through a two-terminal quantum ring
in external magnetic field taking into account the Coulomb interaction with another charge
carrier. The second particle (electron or hole) is confined within the dot settled in the center of the ring.
We assume that the barrier between the ring and the dot is thick enough to neglect the tunnel coupling.
For that confinement potential
model, we perform time evolution of two-particle wave function by solving suitable time-dependent
Schr\"odinger equation. We observe AB oscillations in the electron transfer probability. We also find
that the Coulomb correlation modifies the AB effect in the following way: (i) the maxima of transmission
probability grow when transferred electron is attracted by the charged dot while repulsive interaction lowers
them and (ii) the probability of electron transfer may grow for $(n+1/2)$ magnetic flux quanta piercing the
ring when the interaction is strong enough to excite the carrier that is confined in the inner  dot.
In the latter case, electron transfers part of its energy to the dot. We find  that  the energy
transfer depends on magnetic field due to both the AB effect and the Lorentz force. The electrostatic interaction causes
also positive feedback between transferred electron and the second particle. Even small oscillation of charge in
the dot can perturb potential felt by transferred electron which may change the
phase of the electron wave function in both ring arms. Finally, an inelastic scattering of electron on
oscillating Coulomb potential leads to suppression of AB effect.

The paper is organized in the following way. We define the confinement potential of the considered
system and present our theoretical model in Sec.\ref{sec:theory}. Effect of the repulsive as well as effect
of the attractive interaction on transmission probability are presented in Sec. \ref{sec:ee} and
in Sec.\ref{sec:eh} respectively. Inelastic scattering of the transferred electron on Coulomb potential
in two-terminal quantum ring is analyzed in Sec.\ref{sec:inelastic}. Discussion and conclusions are
provided in Sec.\ref{sec:con}.

\section{\label{sec:theory} Theory}

\begin{figure}[htb!]
\hbox{
	   \epsfxsize=80mm
           \epsfbox[10 220 600 560] {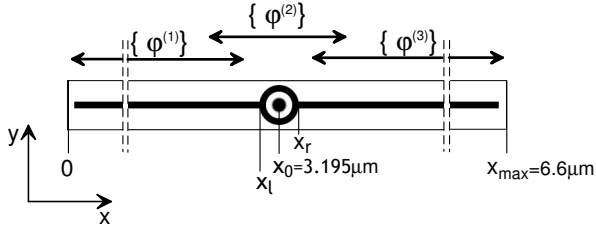}
          \hfill}
\caption{Confinement potential of two-terminal quantum ring with quantum dot placed inside. Arrows
indicate spatial limits of three single-electron wave functions bases used for simulation of
electron wave packet in the leads and in the ring (see text below). Labels $x_{l}$ and $x_{r}$ mark
the left and the right limits of the ring.}
\label{Fig:pot_uw}
\end{figure}

Our confinement potential model consists of quantum ring connected to the left and to the right
leads of finite length and a closed circular quantum dot placed in the center of the ring. Tunneling between the
ring and the dot is neglected due to wide barrier so that the first particle (electron) can only
move in the leads and in the ring while the second (electron or hole) can not leave the dot. The whole
system is put in homogeneous magnetic field which is perpendicular to the quantum ring plane.
The confinement potential is schematically depicted on Fig.\ref{Fig:pot_uw}.
We assume that the confinement is much stronger in the growth (z) direction than that in (x-y) plane
of the ring. Both particles occupy a frozen ground state of the quantization in the growth direction.
System of two interacting particles can be described within an effective two-dimensional
model. We define the confinement potential in the ring ($V_{r}$), in the leads ($V_{l}$) and in the
dot ($V_{d}$) as:
\begin{eqnarray}
 V_{r}(\mathbf{r})&=&V_{e}\exp \left(
-\frac{\left|\left|\mathbf{r}-\mathbf{r}_{0}\right|-r_{r}\right|^{p}}{\sigma_{r}
^{p}}
\right)
\label{eq:potring}\\
 V_{l}(\mathbf{r})&=&V_{e}\exp\left(-\frac{\left|y\right|^{p}}{\sigma_{r}^{p}}
\right)
\label{eq:potkan}\\
V_{d}(\mathbf{r})&=&V_{e(h)}\exp \left(
-\frac{\left| \mathbf{r}-\mathbf{r}_{0}\right|^{p}}{\sigma_{d}^{p}} \right)
\label{eq:potdot}
\end{eqnarray}
In the above equations $V_{e(h)}$ is the maximal depth of the potential for electron (hole),
$\mathbf{r}_{0}$ is the center position of the ring and the dot, $\sigma_{d}$ is the radius of the dot,
$\sigma_{r}$ is the width of the ring arms and both leads as well, $r_{r}$ is the radius of the ring. The
value
of parameter $p$ defines the smoothness of the quantum dot wall. In the calculation we use the following
values: $V_{e}=-200\,\textrm{meV}$, $V_{h}=-140\,\textrm{meV}$,
$\mathbf{r}_{0}=[3.195\,\mu\textrm{m},0]$,
$\sigma_{d}=55\,\textrm{nm}$, $p=8$, $\sigma_{r}=25\,\textrm{nm}$ and $r_{r}=130\,\textrm{nm}$. The
length of the left lead is equal to $3\,\mu\textrm{m}$ while the length of the right lead is
$3.2\,\mu\textrm{m}$.

The main aim of this work is the investigation of the role of Coulomb correlation in the
single-electron transport through the ring. For this purpose we perform the time evolution of
two-particle wave function which fulfills the Schr\"odinger equation:
\begin{equation}
i\hbar\frac{\partial}{\partial t}
|\Psi(\mathbf{r}_{1},\mathbf{r}_{2},t)\rangle
=\widehat{H}|\Psi(\mathbf{r}_{1},\mathbf{r}_{2},t)\rangle
\label{eq:schrodinger}
\end{equation}
where the two-particle Hamiltonian is defined as:
\begin{equation}
\widehat{H}=\widehat{h}_{1}+\widehat{h}_{2}+\frac{q_{1}q_{2}}{4\pi
\epsilon \epsilon_{0}r_{12}}
\label{eq:ham12}
\end{equation}
The Hamiltonians $\widehat{h}_{1}$ and $\widehat{h}_{2}$ are the single-particle
energy operators. The third term on the right hand side of  above equation
describes the electrostatic interaction between the particles which introduces
spatial correlations of their mutual positions. We use single-particle Hamiltonians in the following form:
\begin{equation}
 \widehat{h}_{i}=\frac{(\widehat{\mathbf{p}}_{i}-q_{i}\mathbf{A}(\mathbf{r}_{i}
))^{2}}{2m_{i}^{*}}+V_{o(d)}(\mathbf{r}_{i})
\label{eq:ham1}
\end{equation}
where $\widehat{\mathbf{p}}_{i}=-i\hbar\nabla_{\mathbf{r}_{i}}$ is the particle momentum operator, $m_{i}^{*}$
is effective mass, $q_{i}$ is the charge of the particle, $\mathbf{A}(\mathbf{r})$ is a vector
potential,
$V_{d}(\mathbf{r}_{i})$ is a confinement potential of the dot while
$V_{o}(\mathbf{r}_{i})=V_{r}(\mathbf{r}_{i})+V_{l}(\mathbf{r}_{i})$ is a sum of confinement potential of the
ring and the leads. Since the tunnel coupling between the ring and the dot is neglected in our theoretical
model, the particles confined in spatially separated regions can not exchange their spins. In other words, the
exchange interaction between the electron in the ring and the particle confined in the dot exactly vanishes.
Therefore, all the effects due to the presence of the charged dot inside the ring are only result from the
Coulomb coupling. For non-negligible tunnel coupling between the ring and the dot the single-particle wave
functions of the ring and the dot would overlap. In this case
the exchange correlation  would lead to a dependence of the transmission probability on the relative spin arrangements.
Moreover, for nonzero overlap between the ring and the dot wave functions the
particle confined in the dot might be able to tunnel out to the ring.

The
Hamiltonian (\ref{eq:ham12}) does not depend on the spin coordinates. According to the
superposition principle, we expand the correlated wave function of two spinless particles as linear
combination:
\begin{equation}
 \Psi(\mathbf{r}_{1},\mathbf{r}_{2},t)=
\sum_{i}^{M} c_{i}(t) \psi_{i}(\mathbf{r}_{1},\mathbf{r}_{2})
\label{eq:funkcja2}
\end{equation}
where $c_{i}(t)$ are the time-dependent coefficients and M is the size of the two-particle wave functions
base. The elements $\psi_{i}$ are expressed as products of single-particle wave functions:
\begin{equation}
\psi_{i}(\mathbf{r}_{1},\mathbf{r}_{2})=\varphi_{k(i)}(\mathbf{r}_{1})
\phi_{m(i)}(\mathbf{r}_{2})
\label{eq:baza2}
\end{equation}
where every index \textit{i} corresponds to a particular combination of indices \textit{k} and
\textit{m}. The
\textit{k} index numbers the states of  the first particle which moves in the ring and in the leads
while
\textit{m} numbers the states of the second particle in the quantum dot. In order to find the wave
functions
$\varphi_{k}$ and $\phi_{m}$ we first express them as linear combinations of centered Gaussian functions. For
example, the \textit{k}-th quantum dot state can be written as:
\begin{equation}
\phi_{k}(\mathbf{r})=\sum_{i}a_{i}^{k}
\exp\bigg(
-\frac{(\mathbf{r}-\mathbf{r}_{i})^{2}}{2\sigma^{2}_{g}}
-\frac{iqB}{2\hbar}(x-x_{i})y_{i}
\bigg)
\label{eq:bazagauss}
\end{equation}
where $a_{i}^{k}$ are the linear combination coefficients, $\mathbf{r}_{i}=[x_{i},y_{i}]$ are position vectors
of Gaussian centers, B is the value of magnetic field and q is the charge of the particle ($+e$ for the hole
and $-e$ for the electron). The nodes $\mathbf{r}_{i}$ form a two-dimensional square mesh with the distance
$\Delta_{g}=\sqrt{2}\sigma_{g}$ between neighboring nodes. In next step, we diagonalize the single-particle
Hamiltonians (\ref{eq:ham1}) in the Gaussian functions base (\ref{eq:bazagauss}) in order to find the
coefficients $a_{i}^{k}$. In
calculations we used material parameters for GaAs i.e. effective electron mass $m_{e}^{*}=0.067m_{0}$,
effective heavy hole mass $m_{h}^{*}=0.5m_{0}$ ($m_{0}$ is bare electron mass) and dielectric constant
$\epsilon=12.9$. We used nonsymmetric gauge for the vector potential $\mathbf{A}(\mathbf{r})=B[-y,0,0]$ for which
the magnetic field vector $\vec{B}$ is parallel to the z-axis (perpendicular to the ring plane). The value of
parameter $\sigma_{g}$ was estimated variationally and was equal to $\sigma_{g}=5.16\,\textrm{nm}$.

We determine the single-particle states in the dot by diagonalizing the single-particle Hamiltonian with
confinement potential given by Eq.\ref{eq:potdot}. Therefore,  the wave functions $\phi_{m}$ were determined
only in the quantum dot and in the close surroundings i.e. in the barrier which separates the dot from the
ring. Due to extremely large span of the external subsystem (leads and ring) we divided it into three
overlapping parts. In every spatial parts another single-electron wave functions base is introduced i.e.
$\{\varphi^{(1)} \}$, $\{\varphi^{(2)} \}$ and $\{\varphi^{(3)} \}$ as shown on Figure \ref{Fig:pot_uw}. We
find elements of these three bases in a similar way to the one applied for the quantum dot i.e. diagonalizing the Hamiltonian
(\ref{eq:ham1}) for confinement potential $V_{o}=V_{r}+V_{l}$.
For the preparation of the basis we assumed a different external potential $V_o$ in the three considered regions.
In order to determine the basis elements in a region we modified the potential
assuming $V_o=0$ outside this region in order to spatially limit the basis wave functions  for each region.
As the first second and third region (bases $\{\varphi^{(1)} \}$, $\{\varphi^{(2)} \}$, $\{\varphi^{(3)} \}$) we take  $0<x<2865\,\textrm{nm}$,
2765 nm $<x<$ 3625 nm, and 3525 nm $<x <$ 6600 nm, respectively (see
Fig.\ref{Fig:pot_uw}).

Wave functions
$\{\varphi^{(1)}\}$ and $\{\varphi^{(3)}\}$ are defined in the left (region 1) and in the right
(region 3) leads
respectively, for the distance larger than $200\,\textrm{nm}$ from an outermost parts of the ring (parameters
$x_{l}$ and $x_{r}$ on Fig.\ref{Fig:pot_uw}). In a similar way, the elements $\{\varphi^{(2)}\}$ are
defined in the
second region which covers the ring (without a dot) with parts of both leads to the distance of
$300\,\textrm{nm}$ from the ring. The ranges of these three regions are schematically marked on a
Fig.\ref{Fig:pot_uw}. Notice that the elements of two adjacent basis overlap e.g. the first with the second as
well as the second with the third on the length equal to  $100\,\textrm{nm}$. We carefully checked that these
connections do not perturb motion of electron in both channels.\cite{length} For construction of the
two-particle wave function (\ref{eq:baza2}), we use the lowest energy states obtained from single-particle
Hamiltonian diagonalization. In calculations we use $N_{d}=20$ dot states and  $N_{1}=140$, $N_{2}=60$ and
$N_{3}=150$ states for bases $\{\varphi^{(1)} \}$, $\{\varphi^{(2)} \}$ and $\{\varphi^{(3)} \}$ respectively.

In the Schr\"odinger equation (\ref{eq:schrodinger}) we substitute for
$|\Psi(\mathbf{r}_{1},\mathbf{r}_{2},t)\rangle$ its expansion (\ref{eq:funkcja2}) and next we multiply both
sides of resulting equation by  $\langle \psi_{k}(\mathbf{r}_{1},\mathbf{r}_{2})|$. We obtain the
following matrix equation: \begin{equation} i\hbar \mathbf{H\dot{c}}= \mathbf{Sc} \label{eq:hkm}
\end{equation}
where $\mathbf{S}$ is the overlap matrix of two-particle wave functions basis elements (\ref{eq:baza2})
defined as $S_{km}=\langle \psi_{k}|\psi_{m} \rangle$ while $\mathbf{H}$ is the matrix of two-particle
Hamiltonian (\ref{eq:ham12}) with elements $H_{km}=\langle \psi_{k}|\widehat{H}|\psi_{m} \rangle$. Details of
calculations of the matrix elements of electrostatic interaction are given in previous work [\cite{chwiej1}].
Determination of  these matrix elements are very time consuming and therefore we were forced to limit the
range of Coulomb interaction in the system. We assume the transferred electron does not interact with the
particle confined in the dot if the distance between its position and the dot center exceeds
$390\,\textrm{nm}$. In other words, when electron moves towards the ring it may be partly reflected from a smooth
potential step of height  $\Delta V=0.28\,\textrm{meV}$. The presence of this potential step does not
influence the electron transfer probability since the original kinetic energy of electron on Fermi surface
($E_{F}=1.42\,\textrm{meV}$), considered in this work, is several times larger.\cite{reflection}

The equation (\ref{eq:hkm}) can be numerically solved by using an iterative method, similarly as was shown in
work[\cite{szafran_1el}] for time evolution of electron wave packet in a two-terminal quantum ring. Notice
however, that every iterative method requires very large number of matrix-vector multiplications so that to
retain the stability and to keep the numerical errors as small as possible.  Since, the sizes of matrices
\textbf{H} and \textbf{S} are equal to $7000$, the use of iterative schema in our two-particle problem  would
be inefficient. Instead, we performed the time evolution of two-particle wave function in another
non-iterative way. For this purpose,  we first diagonalized the two-particle Hamiltonian
(\ref{eq:ham12}) and put all obtained
eigenvectors in columns of the new matrix \textbf{U} (of the same size as \textbf{H}). Next, we use this
\textbf{U} matrix to perform the unitary transformation of Eq.\ref{eq:hkm}:
\begin{equation}
i\hbar \bigg(\mathbf{U^{+} SU}\bigg) \bigg(\mathbf{ U^{+}\dot{c}}\bigg)=
\bigg( \mathbf{U^{+}HU}\bigg) \bigg(\mathbf{U^{+}c}\bigg)
\label{eq:hkmu}
\end{equation}
Let us notice that $\mathbf{U^{+}SU}=\mathbf{I}$ where \textbf{I} is the unity matrix and
$\mathbf{U^{+}HU}=\mathbf{D}$ where  \textbf{D} is diagonal matrix with eigenvalues of energy operator
(\ref{eq:ham12}) on a diagonal. Due to the diagonal form of both matrices, the system of $M=7000$ coupled
equations given by Eq.\ref{eq:hkm} transforms into system of M decoupled differential equations:
\begin{equation}
i\hbar  \frac{\partial b_{k}}{\partial t}= D_{kk}b_{k}
\label{eq:bmm}
\end{equation}
where $\mathbf{b}=\mathbf{U^{+}c}$, with solutions:
\begin{equation}
 b_{k}(t)=b_{k}(t=0)\exp\bigg(-\frac{iD_{kk}t}{\hbar} \bigg)
\end{equation}
Obviously in order to obtain solution for original problem defined by Eq.\ref{eq:hkm}, i.e. to obtain values
of coefficients  $c_{k}$, one performs a backward transformation $\mathbf{c}=\mathbf{Ub}$. We made the
diagonalization of \textbf{H} numerically. Therefore, in order to estimate the numerical errors which may
appear due to performing the unitary transformation, we always checked the energy and the norm conservation
for two-particle wave function. The relative errors do not exceed $10^{-6}$.

For $t=0$ we use the following form for the initial two-particle wave function :
\begin{equation}
\Psi_{s}=\Psi(\mathbf{r}_{1},\mathbf{r}_{2},t=0)=\varphi_{0}(\mathbf{r}_{1})
e^{ik_{0}x_{1}}
\phi_{0}(\mathbf{r}_2)
\label{eq:start}
\end{equation}
Wave function $\phi_{0}(\mathbf{r}_{2})$ describes the particle (electron or hole) confined in the dot in the
ground state while $\varphi_{0}(\mathbf{r}_{1})e^{ik_{0}x_{1}}$ is the wave function of the electron moving in
the left channel towards the ring with the average momentum depending on $k_{0}$ value. We determined
$\varphi_{0}(\mathbf{r}_{1})$  by diagonalizing the Hamiltonian (\ref{eq:ham1}) in the centered Gaussian
functions base (\ref{eq:bazagauss}) with the confinement potential: \begin{equation}
V_{s}(\mathbf{r})=V_{l}(\mathbf{r})+\frac{m_{e}^{*}\omega^2}{2}(x-x_{s})^2 \end{equation} where $x_{s}$ is the
center position of harmonic oscillator in the left channel. It was situated in the distance of
$995\,\textrm{nm}$ from the center of the ring (dot). The strength of the harmonic oscillator depends
on the oscillator length ($l_{e}$) i.e. $\hbar\omega=\hbar^{2}/m_{e}^{*}/l_{e}^{2}$. In calculations we used
$l_{e}=50\,\textrm{nm}$. Such way of determination of $\varphi_{0}$ inherently includes the magnetic translation phase change.
For $t=0$ we give the electron in the left channel momentum $\hbar k_{0}$ with $k_{0}=0.05/\textrm{nm}$ which
corresponds to the average energy on the Fermi surface $(E_{F}=1.42\,\textrm{meV})$ in the
two-dimensional electron gas with
density\cite{szafran_1el} $n=4\times10^{10}/\textrm{cm}^{2}$. The choice of initial conditions
i.e. values of parameters such as $ x_{s}$ or $l_{e}$ is quite arbitrary. In Sec.\ref{sec:eh} we will
shortly comment the results obtained also for other sets of initial parameters.

\section{{\label{sec: res}}Results}

Below we denote by $P_{A}$, $P_{B}$
and $P_{C}$ the probabilities of finding the transferred electron in the left channel, within the ring
and in the right channel, respectively. For these quantities we defined auxiliary
operators:
\begin{eqnarray}
\widehat{P}_{A}&=&\Theta(x_{l}-x_{1})\\
\widehat{P}_{B}&=&\Theta(x_{1}-x_{l})+\Theta(x_{r}-x_{1})-1\\
\widehat{P}_{C}&=&\Theta(x_{1}-x_{r})
\label{eq:heaviside}
\end{eqnarray}
In the above definitions $\theta(x)$ is the Heaviside function while
$x_{l}=3040\,\textrm{nm}$ and
$x_{r}=3350\,\textrm{nm}$ are the left and the right limits of the ring in x
direction respectively,
as shown on Fig.\ref{Fig:pot_uw}. Each $P_{i}$ can be simply computed at any
time as expectation
value of specific operator $\widehat{P}_{i}$ i.e. $P_{i}(t)=\langle
\Psi(\mathbf{r}_{1},\mathbf{r}_{2},t) |\widehat{P}_{i}|
\Psi(\mathbf{r}_{1},\mathbf{r}_{2},t)\rangle$ (for $i=A,B,C$). We treat $P_{C}$ and $P_{A}$ are
lower bounds for probability of electron transfer and backscattering, respectively since
the ring is not completely empty at the end of simulations. A part (less than 5\%) of the packet always stays
inside the ring since the sizes of the channels are limited. 

\subsection{{\label{sec:single_el}}Electron transfer without interaction}

\begin{figure*}[htbp!]
\hbox{
	   \epsfxsize=140mm
           \epsfbox {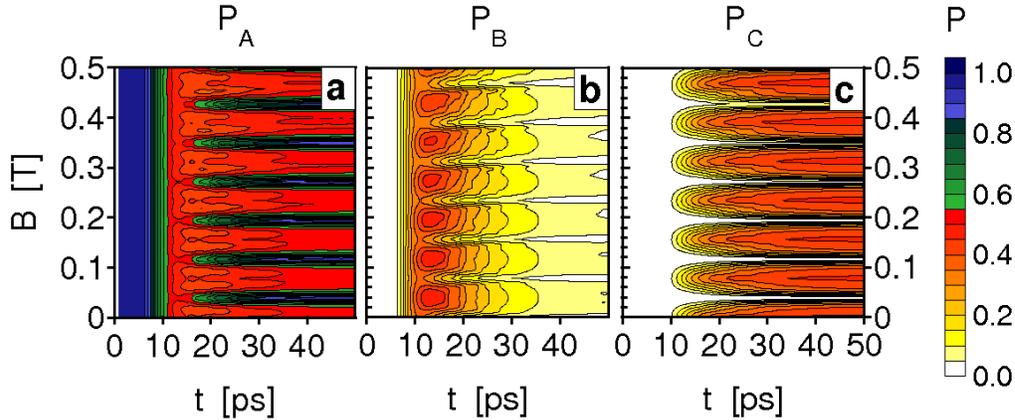}
          \hfill}
\caption{(Color online)
Probabilities  $P_{A}$,  $P_{B}$ and $P_{C}$ as functions of time and magnetic field.
Coulomb interaction between transferred electron and charged dot is neglected.}
\label{Fig:1e_praw}
\end{figure*}

We start the presentation 
by the case when the transferred electron does not
interact with the charged dot. These results will serve as the reference point for the main calculation
where the interactions are included. 
Electrostatic interaction was turned off simply by extracting its matrix elements from
two-particle Hamiltonian (\ref{eq:hkm}). 
The probability distributions $P_{A}$, $P_{B}$ and
$P_{C}$  as functions of time and magnetic field  for this case are depicted on
Fig.\ref{Fig:1e_praw}.

\begin{figure}[htbp!]
\hbox{
	   \epsfxsize=80mm
           \epsfbox[135 131 495 311] {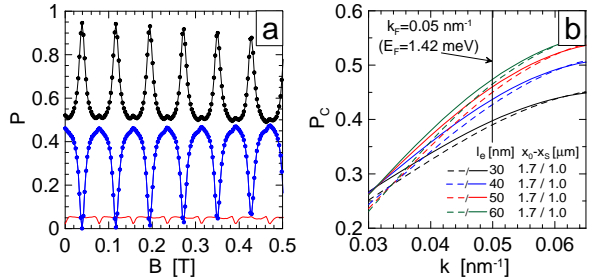}
          \hfill}
\caption{(Color online) a) Probabilities $P_{A}$ (black), $P_{B}$ (red) and $P_{C}$  (blue)
as functions of magnetic field for $t=50\,\textrm{ps}$. Elements of bases
$\left\{\varphi_{1}\right\}$ and
$\left\{\varphi_{2}\right\}$ as well as elements of $\left\{\varphi_{2}\right\}$ and
$\left\{\varphi_{3}\right\}$ overlap on the length of $100\,\textrm{nm}$ (solid lines) and
$200\,\textrm{nm}$ (dots). b) Probability of electron transfer as a function of the initial wave
vector $k_{0}$ for several combinations of parameters $l_{e}$ and $x_{s}$ defining the shape of
single-electron wave packet and its center position for $t=0$. In both cases, the electron does not
interact with charged dot.}
\label{Fig:1e_widma}
\end{figure}

	All probabilities strongly oscillate with magnetic field which is a typical manifestation of
Aharonov-Bohm effect. Period of these  oscillations is $\Delta B=78\,\textrm{mT}$. This value is
close to $\Delta B_{T}=77.98\,\textrm{mT}$ obtained for the one-dimensional ring from formula:
\begin{equation} \Delta B_{T}=\frac{h}{e}\frac{1}{\pi r^{2}} \label{eq:deltab} \end{equation} for
ring radius $r=130\,\textrm{nm}$. Especially the most intensive AB pattern is visible for the
probability of electron transfer [see Fig.\ref{Fig:1e_praw}(c)] i.e. distinct maxima for multiple
integers of magnetic field flux quanta ($\phi_{n}=n(h/e)$ with $n=0,1,2,\ldots$) and blockades of
electron transfer in the half way between adjacent maxima. The presented time-magnetic field
characteristics of probabilities clearly show the dynamics of wave packet motion. For the first 6~ps
the most energetic part of the electron wave packet reaches the left entrance to the ring but then it
takes it about  4~ps to get through the ring to the second junction. This is visible as a large
growth of $P_{C}$ value on Fig.\ref{Fig:1e_praw}(c) for $t\approx 10\,\textrm{ps}$. One can also see
that the electron wave packet leaves the ring more quickly when the  $P_{C}$ is close to its maximum
rather than for its minimum. Besides the AB effect, probabilities of finding the electron in the
left and in the right leads depend also on magnetic field due to the Lorentz
force.\cite{szafran_1el} In order to show magnetic field  effect on electron transport, we made the
cross sections of $P_{A}$, $P_{B}$ and $P_{C}$ distributions shown on Fig.\ref{Fig:1e_praw} for
$t=50\,\textrm{ps}$. These cross sections are shown on Fig.\ref{Fig:1e_widma}(a).
One may notice that the electron transfer through the ring is completely blocked due to AB effect
only in low magnetic field. For example for $B=39\,\textrm{mT}$ probability of an electron transfer
is of the order $10^{-4}$. However, for high magnetic field, the probability of electron transfer
does not drop to zero at all. It means that the AB effect is perturbed by the Lorentz force. Due to
the narrow cross sections of leads and arms of the ring there are no significant changes in the
maxima of transmission probability  as it was theoretically predicted\cite{szafran_1el} and
experimentally observed \cite{3lorentz} for rings with wider arms.

Figure \ref{Fig:1e_widma}(a) shows also comparison of results obtained for $100\,\textrm{nm}$ and
for $200\,\textrm{nm}$ wide overlap regions. Probabilities $P_{A}$ and $P_{C}$  are the same what
proves that electron may smoothly move between neighboring regions without reflection.
	In order to check the influence of initial conditions on the probability of electron
transfer we made additional simulations for several different values of initial parameters that is
for  $k_{0}$, $l_{e}$ and $x_{s}$. Results are presented on Fig.\ref{Fig:1e_widma}(b). We see
that the probability of electron transfer strongly depends on the spatial spread of original wave
packet and its initial momentum rather than its distance from the ring. When initial wave packet becomes
wider (value of $l_{e}$ is larger) then the probability of electron transfer grows even
by several percents. On the other hand, transmission probability is less susceptible for change in
the distance between initial position of wave packet and center position of the ring. Results
obtained for $1.7\,\mu\textrm{m}$ and $1\,\mu\textrm{m}$ are very much the same i.e. the difference
is only about $1\%$.

\subsection{{\label{sec:ee}} Effect of repulsive interaction on electron
transport }

\begin{figure*}[ht!]
\hbox{
	\epsfxsize=150mm
          \epsfbox{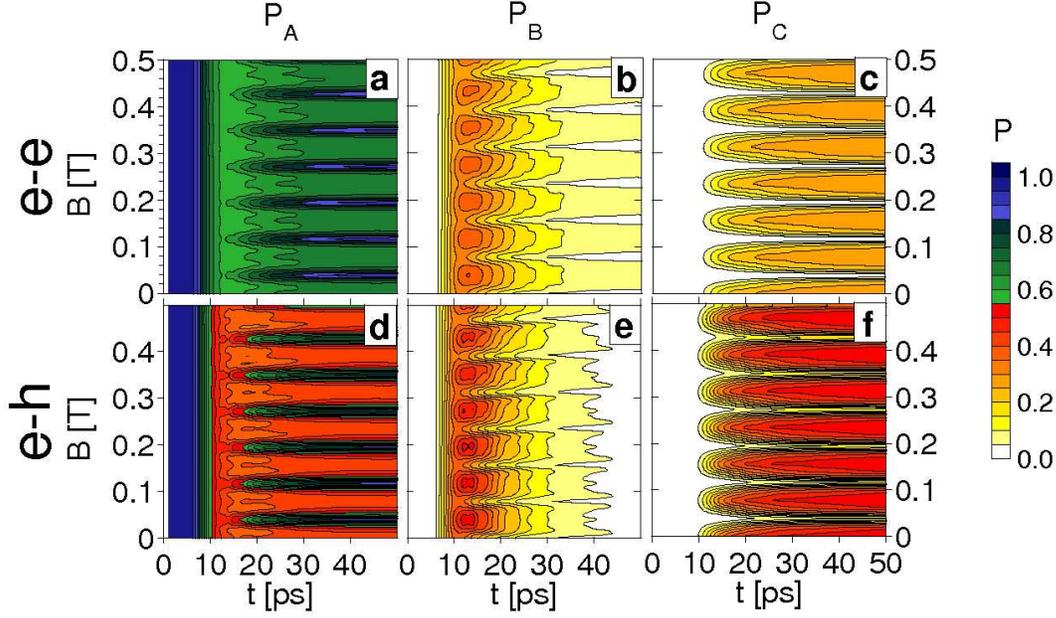}
	\hfill}

\caption{(Color online)
Probabilities $P_{A}$ , $P_{B}$,  $P_{C}$ as functions of time and
magnetic
field for the case when the transferred electron electrostatically interacts with
charged dot.
The figures (a-c) were obtained for repulsive interaction (electron confined in the dot) while
figures (d-f) for attractive interaction (heavy hole confined in the dot).}
\label{Fig:ee_eh_p}
\end{figure*}

\begin{figure}[ht!]
\hbox{
	\epsfxsize=80mm
          \epsfbox{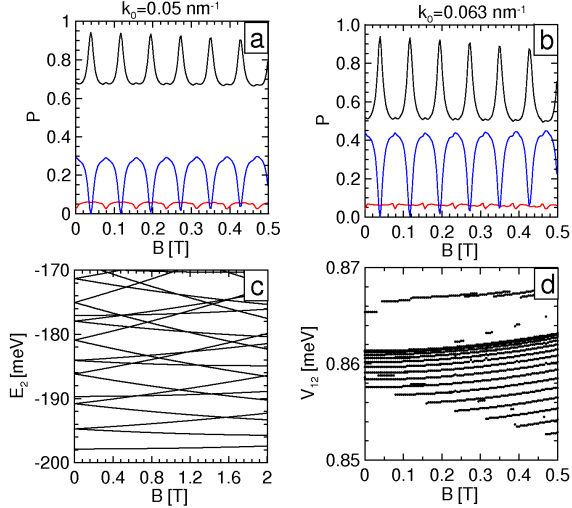}
	\hfill}
\caption{(Color online)
Probabilities $P_{A}$ (black line), $P_{B}$ (red line), $P_{C}$ (blue line)
as functions of magnetic field for $t=50\,\textrm{ps}$ and for (a) $k_{0}=0.05nm^{-1}$ and
(b) $k_{0}=0.063nm^{-1}$. The
transferred electron interacts electrostatically with negatively charged dot.
c) Energy spectrum of single electron confined in the dot.
d) Interaction energies for the lowest-energy states of two electrons confined in closed quantum
ring-quantum dot system .}
\label{Fig:ee_widma}
\end{figure}

        In order to investigate the correlation effects which are due to the repulsive interaction,
we put single electron into the dot and turned on the interaction in the system. Transferred
electron feels a growing repulsive electrostatic potential as it approaches the ring.
Probabilities $P_{A}$, $P_{B}$,$P_{C}$ as functions of evolution time and magnetic field
obtained for this two-electron system are shown on Fig.\ref{Fig:ee_eh_p}. Comparison of
probabilities distributions obtained for electron subject to the repulsive interaction
[Fig.\ref{Fig:ee_eh_p}(a-c)] with those obtained for noninteracting electrons
[Fig.\ref{Fig:1e_praw}] allows us to distinguish
several differences between these two cases. Maxima of transmission probability are decreased for
repulsive interaction in relation to the previous case. Consequently, the interaction is also responsible
for the growth of probability of finding the electron in left lead $P_{A}$
[cf. Figs. \ref{Fig:1e_praw}(a) and \ref{Fig:ee_eh_p}(a)] and also for faster electron wave
packet leakage from the ring for $t>40\,\textrm{ps}$. However, the interaction does not change the
period of AB oscillation. The probabilities $P_{C}$ shown on Fig.\ref{Fig:1e_praw}(c) and on
Fig.\ref{Fig:ee_eh_p}(c) change with the same frequency. For quantitative analysis of interaction
influence on $P_{A}$, $P_{B}$ and $P_{C}$  we have made the cross sections of
probabilities distributions for $t=50\,\textrm{ps}$. These cross sections are presented on
Fig.\ref{Fig:ee_widma}(a).

\begin{figure*}[ht!]
\hbox{
	\epsfxsize=150mm
          \epsfbox{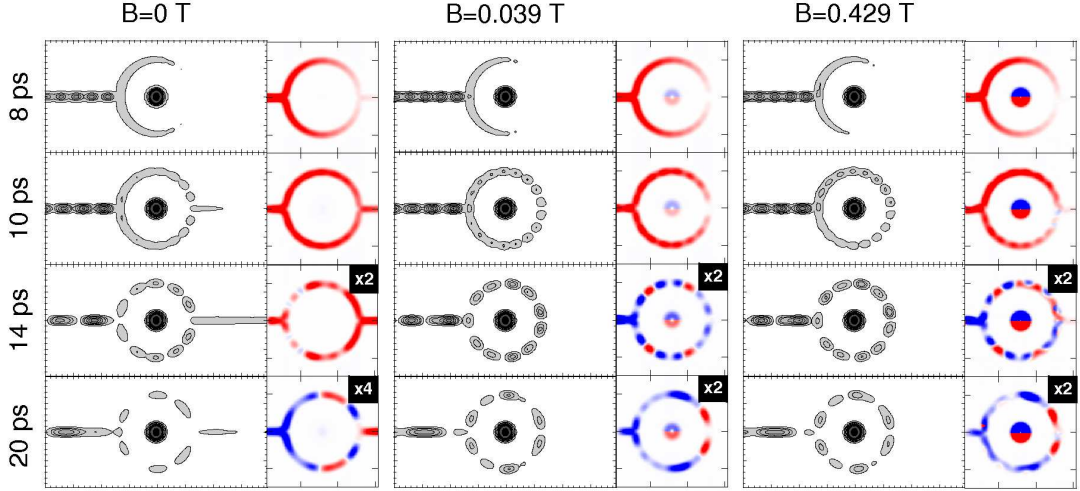}
	\hfill}
\caption{(Color online)
Two-electron probability density (odd columns) and current density (even columns) calculated for
$t=8,10,14,20\,\textrm{ps}$. The red color indicates the current flowing to the right lead while the blue color
marks the current flowing to the left. Intensity of the colors is proportional to the amplitude of
current. The color scales for two lowest rows are enhanced by the multipliers which are shown in
top right corners.}
\label{Fig:eegj}
\end{figure*}

Comparison of $P_{C}$ cross sections shown on Figs.  \ref{Fig:1e_widma}(a) and \ref{Fig:ee_widma}(a)
reveals that repulsive interaction is responsible for about $10\%$ decrease in
transmission probability. It results from the fact that when electron approaches the ring, it
simultaneously climbs on a growing slope of the Coulomb potential of the second electron
and converts part of its kinetic energy into potential energy. Therefore the electron wave packet
enters the ring with lower average wave vector $k$ than its initial value $k_{0}$. Since the
probability of electron transfer strongly depends on the k value [see Fig.\ref{Fig:1e_widma}(b)], the
lower average k value bring the transmission probability down. Such situation is clearly visible on
Fig.\ref{Fig:1e_widma}(b) for $k_{0}<k_{F}$. In order to check this hypothesis we performed
additional time evolution of two-electron wave function giving the transferred electron higher
initial momentum just enough to overcome the repulsive interaction [see Fig.\ref{Fig:ee_widma}(b)].
We assumed $0.86\, meV$ as average value of interaction energy what gives initial momentum $k_{0}=0.063\,
\textrm{nm}^{-1}$ and corresponds to $E_{F}=2.28 meV$.  Probabilities $P_{A}$, $P_{B}$, $P_{C}$
as functions of magnetic field for this case for $t=50\,\textrm{ps}$ are presented on
Fig.\ref{Fig:ee_widma}(b). We notice that this picture is almost identical with results obtained
for electron transport without interaction  [cf. Figs. \ref{Fig:1e_widma}(a) and
\ref{Fig:ee_widma}(b)].
As one may notice, the repulsive interaction does not change the frequency of AB oscillations [cf.
Fig.\ref{Fig:1e_widma}(a) and Fig.\ref{Fig:ee_widma}(b) with Fig.\ref{Fig:ee_widma}(a)]. This
results from the fact that we did not include the term describing interaction between the magnetic
dipole moments in the two-particle Hamiltonian (\ref{eq:ham12}). In addition, the electrostatic
potential originated from charged dot is too weak to induce the electron density redistribution along
the ring radius and thus do not change the effective ring radius [see Eq.\ref{eq:deltab}]. In order
to get a deeper insight into the dynamics of the two-electron wave packet we have calculated the total
two-electron probability densities and the current densities. Results obtained for $t=8,10,14,20
\,\textrm{ps}$ are shown on Fig.\ref{Fig:eegj}.
When the magnetic field is absent in the system, the total electronic density as well as the
current remain symmetrical relative to  $y\to -y$ reflection during the whole time evolution.
Obviously, it results from the fact that the electrostatic interaction term  in two-particle
Hamiltonian (\ref{eq:ham12}) preserves this symmetry and thus does not change the symmetry of
the two-particle wave function. A detailed analysis of the currents for $B=0$ reveals that the
electrostatic interaction does not induce current inside a dot as one at first may  expect. When
electron approaches the ring, both electrons repel each other into opposite
directions. As we do not see any current induced in the dot for $B=0$, we can state that the
electron confined in the dot does not react to the presence of the first electron. This lack of
reaction of the second electron stems from the fact
that the interaction is small in comparison with the lowest single-electron energy excitations in
the dot.
We see on Fig.\ref{Fig:ee_widma}(d) that the value of interaction energy between two electrons
confined in closed quantum ring-quantum dot structure is about $0.86\,\textrm{meV}$. On the
other hand, the energy spacings between the first two excited states and the ground state in quantum
dot confining single electron [see Fig.\ref{Fig:ee_widma}(c)] are equal to $3.1\,\textrm{meV}$ for
$B=0$. The interaction energy is more than three times smaller than even the lowest two
single-electron
energy excitations and therefore can hardly mix the quantum dot states. Since the transferred
electron can not excite the second electron, there is no
energy transfer to the dot. Transferred electron scatters only  elastically on the statical
repulsive potential created in the leads and in the ring by second electron which is confined in the
inner dot. The magnetic field breaks the symmetry of the confinement potential and favors an upper
arm. The larger part of electron wave packet is directed to this arm [see
Fig.\ref{Fig:eegj} for $B=0.039\textrm{T}$ and $B=0.429\textrm{T}$]. Let us notice that the current
in the dot is more intensive for stronger magnetic field. Since the electrostatic interaction is too
weak to induce it, only the external magnetic field may be responsible for its existence. We
explain it by analyzing the matrix elements of probability current:
\begin{equation}
\mathbf{j}_{km}=
\frac{i\hbar}{2m^{*}}
\bigg( \phi_{m}\nabla \phi_{k}^{*}- \phi_{k}^{*}\nabla \phi_{m} \bigg)
-\frac{q}{m^{*}}\mathbf{A}\phi^{*}_{k}\phi_{m}
\label{eq:prad}
\end{equation}
	The first component on the right hand side in Eq.\ref{eq:prad} is the paramagnetic part
of current while the second component is diamagnetic. Now, if we notice that electron
occupies exclusively the ground state of s-symmetry  for B=0 we see that the paramagnetic
current completely disappears.\cite{chwiej2} One may notice on
Fig.\ref{Fig:ee_widma}(c) that even for $B=0.5T$ the energy spacings between the first excited state
and the ground state ($E_{1}-E_{0}=2.7\,\textrm{meV}$) are still much larger than interaction energy.
Since the interaction is not able to mix the dot states, the electron confined in the dot still
occupies the ground state and there is no paramagnetic contribution to the current even in high
magnetic field.
Since the diamagnetic current depends on product of probability density and magnetic field, its
contribution increases for stronger magnetic field what is clearly visible when comparing dot
currents depicted in the fourth and in the sixth columns on Fig.\ref{Fig:eegj}.

\subsection{{\label{sec:eh}}Effect of attractive interaction on electron
transport }

\begin{figure*}[htb!]
\hbox{
          \epsfxsize=140mm
          \epsfbox[125 121 590 268] {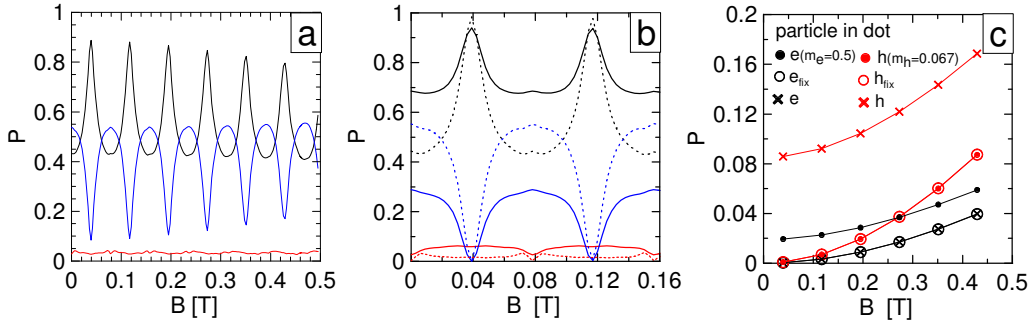}
         \hfill}
\caption{(Color online)
Probabilities $P_{A}$ (black),  $P_{B}$ (red) and  $P_{C}$ (blue) as functions of magnetic field
calculated for a system: (a) with positively charged dot and (b) with electron or hole frozen in the
dot ground state (solid line for electron and dotted one for hole). c) Effect of Coulomb
correlation in the dot on probability of electron transfer for $B=(n+\frac{1}{2})\Delta B$ with
integer n. In (c) the dot is occupied by electron (black color) or by hole (red
color). Lines are guide to the eye.}
\label{Fig:eh_widma}
\end{figure*}
	In the preceding section we showed that the repulsive interaction is responsible for
decrease in probability of electron transfer through the ring. When electron approaches the ring and
negatively charged dot, it is scattered elastically on a static potential. A part of the electron
kinetic energy is converted into the potential energy. The average momentum of the packet is decreased which
leads [Fig.\ref{Fig:1e_widma}(b)] to a decrease in probability of electron transfer. Let us
notice that this mechanism may presumably lead to an increase in the transmission probability provided that
the electron is attracted by the positively charged dot. In order to check this conjecture we put
the heavy hole in the dot and made time evolution of wave function for this electron-hole system.
Probabilities $P_{A}$, $P_{B}$,$P_{C}$ distributions as functions of evolution time and magnetic
field obtained for this case
are shown on Fig.\ref{Fig:ee_eh_p}(d-f). Comparison of the results obtained for the repulsive
interaction [Fig.\ref{Fig:ee_eh_p}(a-c)] with those obtained for the attractive one
[Fig.\ref{Fig:ee_eh_p}(d-f)] shows that the probability of electron transfer through the ring is
indeed larger in the latter case. Moreover, when the transferred electron feels the presence
of the positively charged dot it spends less time in the ring. Attractive interaction leads to an increase
of the packet average momentum and velocity. Therefore, the electron traverses the ring in
shorter time than in the case when it is repelled by negatively charged dot. Cross sections of those
probabilities distributions depicted in Fig.\ref{Fig:eh_widma}(a) indicate that the AB oscillations
are independent of electrostatic interaction. Probabilities $P_{A}$ and $P_{C}$ oscillate with the
same frequency as those shown on Fig.\ref{Fig:ee_widma}(a) obtained for repulsive
interaction, the period of AB oscillation is still equal to $\Delta B=78\,\textrm{mT}$.
The electrostatic interaction does not change the frequency of AB oscillations
but may significantly influence the electron transfer  probability provided that the confinement
along the ring radius is strong. The change of character of electrostatic interaction from repulsive
to attractive makes the maxima of probability of electron transfer grow by more than
$20\%$. 
Repulsive or attractive potential changes the wave vector
distribution in the electron wave packet due to its deceleration or acceleration by
the electrostatic potential. As it is clearly visible on Fig.\ref{Fig:1e_widma}(b) such a change in
the average value of electron wave vector should influence, to a large extent, the probability of
the electron transfer. However, when the electron interacts with a positively charged dot, the minima of
the transfer probability at half flux quanta become shallower.
For example, for  $B=39\,\textrm{mT}$, transmission probability
falls only to $8.4\%$ while the electron transfer is completely blocked when the
electron does not interact with particle confined in a dot [Fig.\ref{Fig:1e_widma}(a)] or
is repelled
by a negatively charged dot [Fig.\ref{Fig:ee_widma}(a)]. Since the Lorentz force is negligible for low
magnetic field, this AB blockade weakness stems only from the interaction of electron wave packet with
the positively charged dot. Figure \ref{Fig:eh_widma}(b) shows probabilities obtained for
attractive and repulsive interaction between the transferred electron and the second particle which is
frozen in the ground state in the dot. Electron or hole confined in the dot can not move and thus
we may neglect the correlation effects in the
dot. Despite this fact, the two-particle wave function is still partly correlated since the
transferred electron interacts with charged dot and its behavior depends on the distance from the
dot due to the Coulomb interaction. We see on Fig.\ref{Fig:eh_widma}(b), that
electron can not be transferred through the ring for $\Delta B/2$ independently of
the character of electrostatic interaction. It means that the Coulomb
correlation in the dot is entirely responsible for the weakness of AB blockade for low magnetic field.
Comparison of the results shown on Figs. \ref{Fig:ee_widma}(a) and \ref{Fig:eh_widma}(a)
suggest that the effect of Coulomb correlation on transmission probability also
depends on the effective mass of particle confined in the dot.  We demonstrate this dependence on
Fig.\ref{Fig:eh_widma}(c) for electron (black crosses), frozen electron (black empty circles) and
electron with large effective mass ($m_{e}^{*}=0.5$ - black dots) as well as for hole (red
crosses),
frozen hole (red empty circles) and hole with small effective mass ($m_{h}^{*}=0.067$ - red dots)
confined in the dot. The results for the frozen hole and the hole with a small mass are identical. This shows that small
effective
mass prevents particle from moving inside the dot regardless of the character of electrostatic
interaction. For heavier particle i.e. electron or hole confined in the dot with effective mass of
about $0.5$, probability of electron transfer for $B=39\,\textrm{mT}$ is increased.
However, this growth is bigger for the attractive ($8\%$) than for the repulsive ($2\%$)
interaction.
Notice also that the transmission probability
grows faster for the  attractive interaction than those for the repulsive one.

\begin{figure}[htb!]
\hbox{
          \epsfxsize=80mm
          \epsfbox{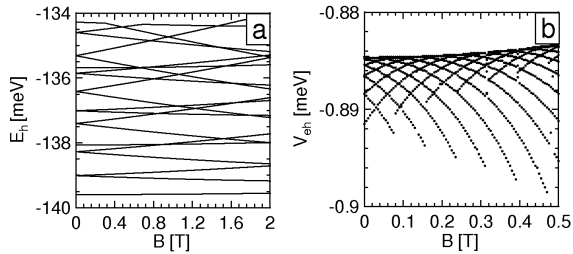}
         \hfill}
\caption{a) Energy spectra of heavy hole confined in the dot and b)
electron-hole interaction energy
in closed quantum ring-quantum dot system.}
\label{Fig:veh}
\end{figure}
	Relatively large effective mass of the heavy hole leads to its stronger localization in the dot.
This results in smaller spacings between the lowest energy levels than those for electron [cf.
Fig.\ref{Fig:veh}(a) for hole and Fig.\ref{Fig:ee_widma}(c) for electron]. For the hole confined in
the dot, these spacings are comparable with the average absolute value of the attractive interaction.
For example, in the absence of magnetic field, the two lowest excited states shown on
Fig.\ref{Fig:veh}(a) lie only $0.59\,\textrm{meV}$ above the ground state while the absolute average
value of interaction energy between electron and hole shown on Fig.\ref{Fig:veh}(b) for closed
quantum dot-quantum ring system is about $0.885\,\textrm{meV}$. Therefore, when the transferred
electron approaches the positively charged dot placed in the center of the ring, it may quite easily
excite the hole in the dot.
\begin{figure*}[ht!]
\hbox{
	\epsfxsize=150mm
          \epsfbox{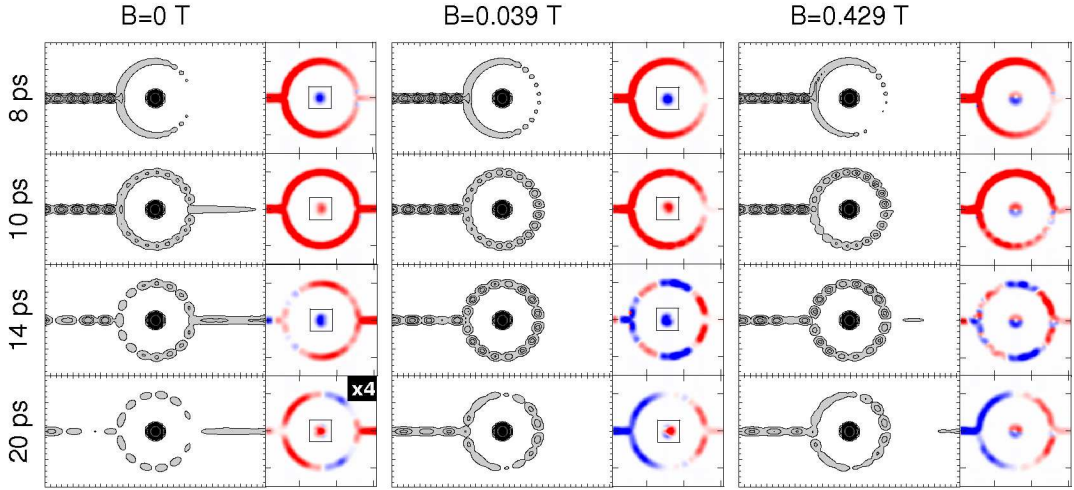}
	\hfill}
\caption{(Color online) The electron-hole probability density (odd columns) and current density
(even columns) for $t=8,10,14,20\,\textrm{ps}$. Red and blue colors indicate directions of current
flow, to the right and to the left respectively. Intensity of colors are proportional to the
amplitudes of currents. The scale for the currents in the dot was enhanced four times (square
region) for $B=0$ and  $B=39\,\textrm{mT}$.}
\label{Fig:ehgj}
\end{figure*}

Figure \ref{Fig:ehgj} shows the total probability density and current density distributions obtained for
fully correlated electron-heavy hole system. For $B=0$ and $t=8\,\textrm{ps}$, when the electron
enters the ring through the left junction, the hole is attracted by the electron and starts to move which
induces the current in the dot. At first, this current flows to the left (blue color on
Fig.\ref{Fig:ehgj}), but when the electron fills more or less equally the ring ($t=10\,\textrm{ps}$)
which makes the potential in the dot less perturbed, the hole reflects from the wall. Then the
current within the dot flows to the right (red color). The hole is excited in the dot and starts to
oscillate. Its spatial oscillation do not fade out even for long time e.g.
$t=20\,\textrm{ps}$. This indicates that when the electron passes through the ring, it transfers
part of its energy to the dot. As the energy of the electron changes permanently, we may state that  it
scatters inelastically on the Coulomb potential generated by the oscillating
hole. Similar oscillations of current in the dot in the horizontal direction are also visible for
$B=39\,\textrm{mT}$  (fourth column on Fig.\ref{Fig:ehgj}).
We will analyze in detail this process of energy transfer between
the electron and the dot in next section.

Horizontal oscillations of the hole in the dot perturb the potential in both arms of the ring.
Although, the confinement potential of the ring is perturbed, it remains symmetrical relative to
$y\to -y$ reflection. That produces identical phase shifts in both parts of electron wave packet i.e.
in the upper and in the lower ring
arms. In other words, the weakness of AB blockade observed on Fig.\ref{Fig:eh_widma}(a) is not a
result of dephasing\cite{mohanty} because the phases in the upper and in the lower parts of the
electron wave packet still change coherently. In consequence, when they meet at the second junction
for $B=(n+1/2)\Delta B$, their phase difference is no longer equal to $\pi$ due to
potential perturbation. This effect was recently
predicted by Chaves et al.\cite{chaves} They obtained very similar dependence of transmission
probability on magnetic field to that shown on Fig.\ref{Fig:eh_widma}(a) for an open two-dimensional
ring with two static impurities put near both arms of the ring and placed symmetrically to its
center.

For high magnetic field, e.g. $B=0.429\,\textrm{T}$ [the last column on Fig.\ref{Fig:ehgj}], these
current oscillations become invisible and now the current encircles the dot in the clockwise direction
i.e. in the opposite direction to the one of the last column of Fig.\ref{Fig:eegj} when electron
occupies the dot. It does not mean
that the oscillations entirely disappear, but only the diamagnetic contribution to the current in
the dot is much larger than paramagnetic contribution. Such large diamagnetic current  was also
induced by magnetic field when an electron was confined in the dot. However, if we compare the dot
currents in the last columns of Fig.\ref{Fig:eegj} and of Fig.\ref{Fig:ehgj} we will see that the
current is less intensive for the hole (color scales on both figures are the same). To explain this
fact we make an assumption that the densities of electron and hole in the dot do not differ much for
the same magnetic field which seems reasonable for our case, since the confinement potential of the
dot is quite strong. With this assumption, and for fixed value of magnetic field, the absolute
value of diamagnetic term in Eq.\ref{eq:prad} depends only on the effective mass of particle. Since
the diamagnetic current is  inversely proportional to the effective mass and the effective mass of the electron
used in calculation was about $m_{h}/m_{e}=7.5$ times smaller than effective mass of the heavy hole,
the diamagnetic contribution to the current is by about $m_{h}/m_{e}$ larger
for the electron than that for the hole.

\begin{figure}[ht!]
\hbox{
	\epsfxsize=80mm
          \epsfbox[150 398 380 620]{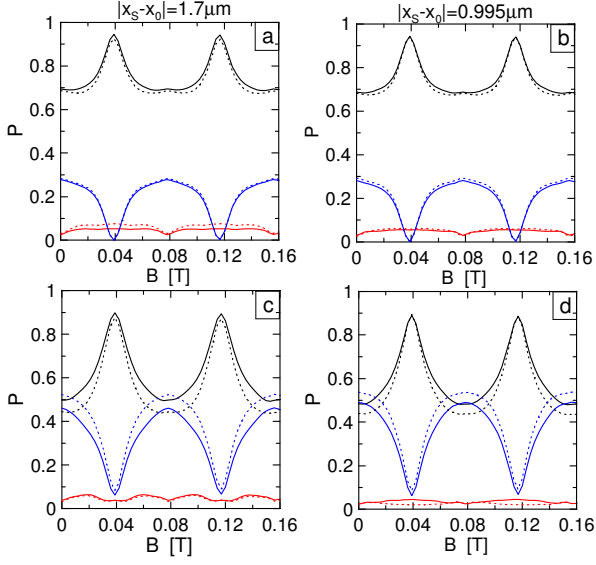}
	\hfill}
\caption{(Color online) The  $P_{A}$ (black), $P_{B}$ (red) and $P_{C}$ (blue) probabilities as
functions of magnetic field for four combinations of $|x_{S}-x_{0}|$ and $l_{e}$ initial parameters.
Results were obtained for $t=50\,\textrm{ps}$. Pictures (a) and (b) are for repulsive interaction
while (c) and (d) are for attractive interaction. Solid lines are for $l_{e}=30\,\textrm{nm}$ and
dotted for $l_{e}=50\,\textrm{nm}$.}
\label{Fig:wp_abcd}
\end{figure}

	Probability of electron transfer depends also on the initial conditions which we have assumed quite arbitrarily.
In order to check the sensitivity of the transmission probability to initial conditions, we studied
the time evolution of the two-particle wave function (\ref{eq:funkcja2}) for four combinations of
the distance between the initial position of electron wave packet and the center position of the ring
($|x_{s}-x_{o}|$) and its spatial span $l_{e}$.
Probabilities $P_{A}$, $P_{B}$ and $P_{C}$ calculated for these new initial parameters
and for $t=50\,\textrm{ps}$ are shown in Fig.\ref{Fig:wp_abcd}(a-d). If the transferred electron
interacts with negatively charged dot, these probabilities are only slightly
sensitive to the change of the initial conditions [see Figs.\ref{Fig:wp_abcd}(a) and
\ref{Fig:wp_abcd}(b)]. For example, transmission probability increases only about $1\%$ when the
parameter $l_{e}$ changes from $30\,\textrm{nm}$ to $50\,\textrm{nm}$ for
$|x_{s}-x_{o}|=0.995\,\mu\textrm{m}$. Much larger differences in transmission probability
were found for system with the positively charged dot.
Generally, the amplitude of AB oscillations is larger for larger $l_{e}$ and when the
electron wave packet stays closer to the ring for $t=0$. For example, for $B=0$ and
$|x_{s}-x_{o}|=1.7\,\mu\textrm{m}$,  the transmission probability grows from about $0.46$ for
$l_{e}=30\,\textrm{nm}$ to about $0.52$ for $l_{e}=50\,\textrm{nm}$ what gives the growth
of about $6\%$ while it is equal to about $4.7\%$ for $|x_{s}-x_{o}|=0.995\,\mu\textrm{m}$.
When the parameter $l_{e}$ is fixed, then the change in $|x_{s}-x_{o}|$ value make less impact on
the transmission probability. For example, for $l_{e}=30\,\textrm{nm}$, we get the increase
in transmission probability of about $3\%$ when the initial position
of the electron wave packet is shifted by about $0.7\,\mu\textrm{m}$ closer to the ring
whereas for
$l_{e}=50\,\textrm{nm}$ the increase in $P_{C}$ value is less distinct and is equal to about $1.1\%$
then. On the other hand, weakness of AB blockade for electron-hole system is independent of the
initial position of transferred electron wave packet but grows by $2\%$ when the value of
parameter $l_{e}$ changes from $30\,\textrm{nm}$ to $50\,\textrm{nm}$ for $B=39\,\textrm{mT}$.

\section{\label{sec:inelastic} Elastic and inelastic scattering}

\begin{figure*}[htb!]
\hbox{
	\epsfxsize=150mm
          \epsfbox[15 416 829 680]{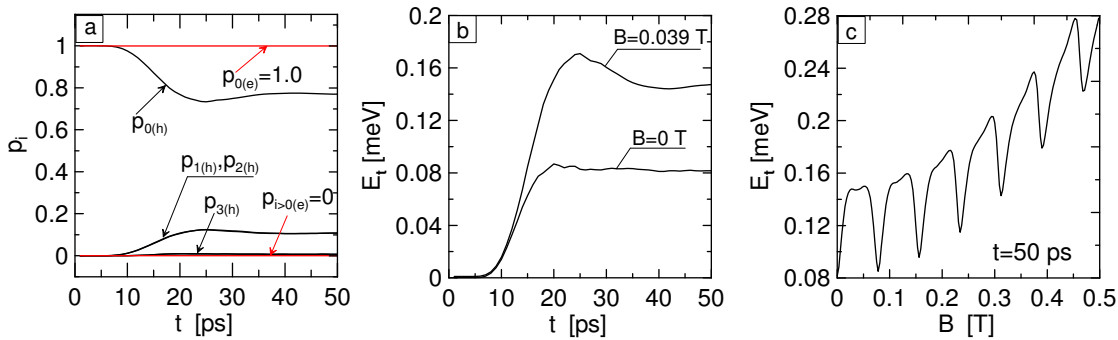}
	\hfill}
\caption{(Color online) a) Probabilities of occupation of the low-energy quantum dot states by
the electron (red) and by the hole (black) as functions of the evolution time. In (a) results were obtained for
$B=39\,\textrm{mT}$. b) Energy gained by the hole confined in the dot as function of time.
 c) Permanent energy transfer to positively charged dot depending on magnetic field.}
\label{Fig:transfer}
\end{figure*}

In the previous section we showed that during the electron transition through the ring, the
particle confined in the dot may start to move. Its spatial oscillations within the dot are induced by the
electrostatic interaction between charged particles and are due to the excitation to the higher
energy states in the dot. During the process of excitation, the transferred electron loses a part of its
kinetic energy which is gained by the second particle. If this energy loss is permanent i.e. the
electron does not recover it after it leaves the ring, then the process of electron scattering on the Coulomb
potential is inelastic. Figure Fig.\ref{Fig:transfer}(a) shows the probabilities of
occupation of the low-energy quantum dot states
as functions of evolution time. This picture was obtained for $B=39\,\textrm{mT}$. In order to find
the probability of
occupation of the particular dot state we have projected the two-particle wave function
(\ref{eq:funkcja2}) on that state:
\begin{equation}
{p}_{i}(t)=
\langle \Psi(\mathbf{r}_{1},\mathbf{r}_{2},t)|\widehat{p}_{i}|
\Psi(\mathbf{r}_{1},\mathbf{r}_{2},t) \rangle
\end{equation}
where $\widehat{p}_{i}=|\phi_{i}\rangle \langle\phi_{i}|$ is the projection
operator.

When electron is confined in the dot the probabilities of occupation of the dot states do not change in time.
For the confinement potential model considered here, the electron is always in the ground
state. As it was mentioned in Sec.\ref{sec:ee} the electrostatic interaction is too weak to excite
the electron within the dot and this is the reason we do not see any current in the dot on
Fig.\ref{Fig:eegj} for $B=0$. Situation changes dramatically if we consider the hole confined in
the dot. We see on Fig.\ref{Fig:transfer}(a) that after a few ps, the hole starts to be excited
since the probabilities of two the lowest excited states with angular momentum $L=1$ grow with time.
Contributions of these states are identical, since their linear combination gives the hole
oscillation in the horizontal direction. Obviously, it results from the symmetry of the
confinement potential model relative to $y \to -y$ reflection and due to the absence of Lorentz force in the
system for such
small magnetic field. Other hole states in the dot remain unoccupied. The process of the hole excitation
ends up for $t=25\,\textrm{ps}$. During the  next $15\,\textrm{ps}$, the hole partly de-excite and
the probability of finding it in the ground state is increased. For $t>40\,\textrm{ps}$,
contributions from the low-energy dot states  stabilize. These changes of probabilities of
occupation of the dot states influences the energy of the hole. We calculated the energy
gained by hole i.e. energy transfer to the dot, from following formula:
\begin{equation}
 E_{t}(t)=\sum_{i=0}^{19} p_{i}(t)  E_{i}^{dot}-E_{0}^{dot}
\end{equation}
where $E_{i}^{dot}$ are the eigenenergies of particle confined in the dot. Figure
\ref{Fig:transfer}(b) shows the time characteristics of the energy transferred to the dot which is
occupied by the hole for $B=0$ and $B=39\,\textrm{mT}$. We see that both cases differ qualitatively
as well as quantitatively. In the absence of magnetic field, the energy is transferred to the dot
for $t<20\,\textrm{ps}$. Then the hole energy changes  only slightly and for $t=50\,\textrm{ps}$
it stabilizes at about $0.08\,\textrm{meV}$. Thus the transferred electron lose $5.6\%$ of its
original kinetic energy. For $B=39\,\textrm{mT}$  the energy transfer in the first $25\,\textrm{ps}$
is twice of that observed for $B=0$. Next, the hole gives back a part of the gained energy to the
electron but for $t=50\,\textrm{ps}$ is still much larger than in the case for $B=0$.
The occurrence of such a distinct difference in energy transfer is not incidental.
The magnetic field dependence of the energy
transferred to the dot occupied by the hole, depicted on Fig.\ref{Fig:transfer}(c), indeed have
minima for $B=n\Delta B$ i.e. for maxima of the transmission probability. On the other hand, the
maxima of the energy transfer do not appear exactly for $B=(n+1/2)\Delta B$, as one may expect, but they
are shifted towards higher magnetic fields. On Fig.\ref{Fig:transfer}(c), we see that, besides
the oscillatory character of magnetic field dependence of energy transfer what is the signature of
AB effect, the minima and maxima of energy gained by the hole lie higher in energy when
magnetic field becomes stronger. This nonlinear effect is the signature of presence of magnetic
force in the system. Magnetic field breaks the symmetry of the confinement potential of the ring and
consequently the Lorentz force injects larger part of the electron wave packet to the upper arm of
the ring  [see density distributions on Fig.\ref{Fig:eegj} and Fig.\ref{Fig:ehgj} for
$B=0.429\,\textrm{T}$ and $t=8\,\textrm{ps}$]. In this case, the magnitude of Coulomb interaction
between both particles is getting stronger. It results in a larger amount of the energy transferred to
the dot. For example, for $B=0.453\,\textrm{T}$ the energy gained by the hole reaches even
$0.278\,\textrm{meV}$ what is $19.5\%$ of original kinetic energy of the transferred electron.
\section{\label{sec:con} Discussion and conclusions}

The presence of a charged dot in the center of the ring significantly influences the probability of
electron transmission. The maxima of transmission probability observed in the Aharonov-Bohm effect are
shifted down (up) for repulsive (attractive) interaction between transferred electron and the
charged dot [cf. Fig.\ref{Fig:1e_widma}(a) for empty dot with Figs. \ref{Fig:ee_widma}(a) and
\ref{Fig:eh_widma}(a)]. The reduction of transmission probability stems from
lowering the average value of
wave vector in the electron wave packet [see Fig.\ref{Fig:1e_widma}(b)] due to
deceleration of its
motion when it moves towards the ring. The magnitude of this probability reduction depends in
particular on the radius of the ring  and on the number of particles confined in the dot.
Interaction should be stronger for smaller rings due to stronger Coulomb coupling of the ring and
the dot, and for multiple charged dot. Moreover, the probability of electron transmission may also
be decreased when the kinetic energy of electron is of the same order as the interaction energy.
Then, the low-energy part of the electron wave packet should be reflected back from repulsive
Coulomb potential before it get closer to the ring.

The single electron transport in quantum ring depends strongly on the relations between the
magnitude of interaction energy and the lowest excitation energies of particle confined in quantum
dot. When the spacings between the two lowest excited states and the ground state in the dot are
several times larger than the interaction energy [cf. Figs. \ref{Fig:ee_widma}(c) and
\ref{Fig:ee_widma}(d)], the electron transport through the ring is blocked for $(n+1/2)\phi_{0}$ flux quanta in
low magnetic field [see the  magnetic field dependence of $P_{C}$ (blue color) on Figs.
\ref{Fig:ee_widma}(a) and \ref{Fig:ee_widma}(b)]. In this case, the transferred electron is not able
to excite the second particle which stays in the ground state [the $p_{i}$ do not change for
repulsive interaction (red color) on Fig.\ref{Fig:transfer}(a)] and the Coulomb potential originated
from charged dot keeps its azimuthal symmetry.  In consequence, the quantum interference is not
perturbed by the interaction because the transmitted electron scatters elastically in quantum ring
i.e. there is no permanent energy transfer between the electron and the charged dot.

The situation changes significantly when the interaction energy becomes comparable to excitation energies.
For example, the magnetic field dependence of transmission probability obtained for attractive
interaction and presented on Fig.\ref{Fig:eh_widma}(a) reveals AB blockade weakness for
$(n+1/2)\phi_{0}$ even for low magnetic field. The heavy hole is excited by the transferred electron
due to their Coulomb interaction [see $p_{i}$ for attractive interaction (black color) on
Fig.\ref{Fig:transfer}(a)], and starts to oscillate horizontally within the dot [see the currents on
Fig.\ref{Fig:ehgj} for $B=0$ and $B=39\,\textrm{mT}$]. Coulomb potential originated from oscillating
hole charge, breaks the azimuthal symmetry of the confinement potential in the ring. This dynamical
charge redistribution inside the dot perturbs the quantum interference in the ring. Electron scatters
inelastically on the oscillating Coulomb potential which changes coherently
the phase of the electron wave packet in both arms of the ring. Finally, this leads to the suppression
of AB effect i.e. the maximum-to-minimum ratio is decreased but the amplitude of transmission
probability does not change much. Interestingly, the energy gained by the charge confined in the dot
shows strong oscillation in the magnetic field [see Fig.\ref{Fig:transfer}(c)]. Maxima are localized in
the proximity
of $(n+1/2)\phi_{0}$ and are slightly shifted towards the higher magnetic fields.

Generally the AB oscillation period depends on: (i) the effective radius of the ring and (ii) the vector
potential. Oscillation of charged particle within the dot creates an additional magnetic field and vector
potential. However this induced magnetic field is very small\cite{chwiej2} i.e. of the order of few hundreds
nT and therefore this effect can not perturb significantly the AB period. The Coulomb interaction may potentially
influence the effective ring radius since the electron tends to move closer to the positively charged dot due
to attractive interaction whereas it tries to keep away from negatively charged dot due to repulsive
interaction when it traverses the ring. However the effective ring radius can be changed only if the
interaction is strong enough to modify the electron density along the ring radius what is possible for very
wide ring arms. For  confinement potential model considered here, the interaction is too weak to make
noticeable redistribution of electron density in the ring and we did not observe any change in AB
period due to the Coulomb interaction.

Similar effect i.e. suppression of AB oscillation in conductance  was observed in experiment of
M\"uhle\cite{ex_ring_ring_1} for two capacitively coupled quantum rings. They obtained much less distinct AB
oscillations for outer ring than the conductance oscillation arising from AB effect for the inner ring. The
authors ascribes this effect to the imperfections of the confinement potential of the outer ring. However it
does not explain such large amplitude of oscillation induced by the inner ring since the charge redistribution
in the inner ring perturbs the Coulomb potential felt by electrons in the outer ring.
In our opinion besides the imperfections of the outer ring, the difference in amplitudes of AB oscillations
observed in experiment  results also from inelastic scattering of the transferred electrons on
the Coulomb potential.
Since the energy gaps between the ground state and the first excited state are much smaller in the ring than in
the dot, the particle confined in the inner ring should be much more easily excited i.e. much energy
may be transferred to the inner ring than to the dot. In such a case, strong spatial oscillations of
particle within the inner ring may govern the motion of the electron injected to the outer
ring. Finally, this strongly inelastic scattering process can suppress the AB
oscillation of the outer ring rather than the amplitude of the electron transmission probability.

In conclusion, the effect of Coulomb correlation on single-electron transport in
two-terminal quantum ring capacitively coupled to the charged dot was
theoretically investigated. The Coulomb interaction between the transferred electron
and charged particle confined in the dot, significantly influences the
maxima of transmission probability in Aharonov-Bohm effect. When
interaction energy is comparable to the lowest excitation energies in
the dot then the electron transfers part of its energy to the dot. Thus
electron scatters inelastically on the ring which finally leads to a reduction of
Aharonov-Bohm blockade and suppression of Aharonov-Bohm oscillation.

\begin{acknowledgements}
We are grateful to B. Szafran for useful discussions.
This work was supported by Polish Ministry of Science and Higher Education
within the project N N202 103938 (2010-2013).
Calculations were performed in
ACK\---CY\-F\-RO\-NET\---AGH on the RackServer Zeus.
\end{acknowledgements}


\begin{thebibliography}{00}
\bibitem{length}We made several simulations for wider overlapping parts i.e.
for $200\,\textrm{nm}$, and did not find any change in results.
\bibitem{reflection} We checked that results obtained for potential step
$\Delta V=0.19\,\textrm{meV}$ (negatively charged dot) when the range of the
Coulomb interaction between particles is limited to $600\,\textrm{nm}$ were
identical with those for $\Delta V=0.28\,\textrm{meV}$. It results from the fact,
that in both cases the same low-energy part of the electron wave packet
reflects due to Coulomb interaction and does not enter the ring at all.
\end{thebibliography}

\end{document}